\documentclass[%
 aip,
 amsmath,amssymb,
 reprint,%
]{revtex4-1}

\usepackage{graphicx}
\usepackage{dcolumn}
\usepackage{bm}

\usepackage[utf8]{inputenc}
\usepackage[T1]{fontenc}
\usepackage{mathptmx}
\usepackage{etoolbox}

\makeatletter
\def\@email#1#2{%
 \endgroup
 \patchcmd{\titleblock@produce}
  {\frontmatter@RRAPformat}
  {\frontmatter@RRAPformat{\produce@RRAP{*#1\href{mailto:#2}{#2}}}\frontmatter@RRAPformat}
  {}{}
}%
\makeatother
\begin{document}

\preprint{AIP/123-QED}

\title{A Multi-Pass Optically Pumped Rubidium Atomic Magnetometer with Free Induction Decay}
\author{Lulu Zhang}

\author{Yongbiao Yang}
\author{Ni Zhao}
\affiliation{ 
State Key Laboratory of Quantum Optics and Quantum Optics Decices, Shanxi University, Institute of Opto-Electronics,Taiyuan 030006, China
}%

\author{Jun He}%
\author{Junmin Wang}
 \homepage{Corresponding author: wwjjmm@sxu.edu.cn; ORCID : 0000-0001-8055-000X}
\affiliation{ 
State Key Laboratory of Quantum Optics and Quantum Optics Decices, Shanxi University, Institute of Opto-Electronics,Taiyuan 030006, China
}%
\affiliation{
Collaborative Innovation Center of Extreme Optics, Shanxi University, Taiyuan 030006, China
}%

\begin{abstract}
A free\mbox{-}induction\mbox{-}decay (FID) type optically\mbox{-}pumped rubidium atomic magnetometer driven by a radio-frequency (RF) magnetic field is presented in this paper. Influences of parameters, such as the temperature of rubidium vapor cell, the power of pump beam, and the strength of RF magnetic field and static magnetic field on the amplitude and the full width at half maximum (FWHM) of the FID signal, have been investigated in the time domain and frequency domain. At the same time, the sensitivities of the magnetometer for the single-pass and the triple-pass probe beam cases have been compared by changing the optical path of the interaction between probe beam and atomic ensemble. Compared with the sensitivity of $\sim$21.2 pT/$\rm{Hz^{1/2}}$ in the case of the single-pass probe beam, the amplitude of FID signal in the case of the triple-pass probe beam has been significantly enhanced, and the sensitivity has been improved to $\sim$13.4 pT/$\rm{Hz^{1/2}}$. The research in this paper provids a reference for the subsequent study of influence of different buffer gas pressure on the FWHM and also a foundation for further improving the sensitivity of FID rubidium atomic magnetometer by employing a~polarization-squeezed light as probe beam, to achieve a sensitivity beyond the photo\mbox{-}shot\mbox{-}noise level. 
\end{abstract}

\maketitle

\section{\label{sec:level1}Introduction}

Magnetometers are important tools for measuring magnetic field in many fields, especially weak magnetic fields. Different types of magnetometers have different sensitive ranges. The minimum measurable magnetic field of fluxgate magnetometer, which is one of the most widely used commercial magnetometers and is based on electromagnetic induction, is generally at~nT level. Nitrogen-vacancy (NV) center in diamond magnetometer$^{1}$, which uses the NV center in diamond, has demonstrated magnetic measurements that have reached pT level, and~it can be used in condensed matter physics and other fields. Superconductor quantum interference device (SQUID) magnetometer$^{2}$, based on ultra\mbox{-}low\mbox{-}temperature liquid helium system, has reached ~sub-fT level, but high maintenance costs and harsh environmental requirements have prevented it from being widely used. An atomic magnetometer can measure magnetic field at aT level, which is comparable to SQUID. Moreover, it is easily miniaturized and its low cost means it  has been developed quickly in recent decades. At present, atomic magnetometers are widely used for magnetocardiography and magnetoencephalography in medical fields$^{3-5}$ and cosmic dark matter detection in space exploration fields$^{6-8}$. It also plays an important role in military exploration, navigation, and~other fields$^{9-11}$. The~types of atomic magnetometers include the spin-exchange relaxation-free (SERF) atomic magnetometer, the~Mz magnetometer, the~Mx magnetometer, and etc. The~SERF atomic magnetometer mainly works at low frequencies and in near zero magnetic environments with very high atomic number densities$^{12}$; its sensitivity has reached the order of sub-fT/$\rm{Hz^{1/2}}$. The~Mz magnetometer$^{13}$ and the Mx magnetometer$^{14}$ generally use a circularly or elliptically polarized beam to polarize atoms and detect a magnetic field. In~addition, the~coherent population trapping (CPT) atomic magnetometer$^{15}$, the~all-optical Bell--Bloom atomic magnetometer$^{16}$, and the free\mbox{-}induction\mbox{-}decay (FID) atomic magnetometer$^{17,18}$ are also widely~used.

FID atomic magnetometer extracts magnetic-field information via the Larmor frequency. In the process of atomic spin polarization, Erling Riis~et~al. used elliptically polarized light in a microfabricated cesium vapor cell to polarize an atomic sample$^{19}$. The effects of frequency modulation and amplitude modulation on sensitivity were compared and they realized a magnetic sensitivity of 3.9 pT/$\rm{Hz^{1/2}}$ by amplitude modulation. S. G. Li~et~al. adopted continuous or burst sampling modes for analyzing the frequency performance of a system$^{20}$. A high measurement bandwidth was obtained whilst maintaining good sensitivity in a continuous sampling mode; this was helpful for further applications in magnetometers or sensor arrays with high measurement bandwidth requirements. H. F. Dong~et~al. demonstrated a distributed magnetic field measurement scheme based on FID magnetometer$^{21}$. The magnetic sensitivity reached 10 pT/$\rm{Hz^{1/2}}$ in a bandwidth of 0.6--100 Hz. This mode could image the magnetic field distribution in the measurement range, which could be used in magnetic source positioning and other applications. In addition to static magnetic field measurements, FID magnetometer can also be used to track and measure a variable magnetic field$^{22-24}$. The multi\mbox{-}channel scheme was also used in magnetic field measurement devices. The~current multi\mbox{-}channel schemes generally included two cases. Characteristically, D. Sheng~et~al. placed two cylindrical mirrors in a rubidium vapor cell to form 42\mbox{-}pass configuration and used two beams in the reverse direction as a gradiometer$^{25}$. A sensitivity of 0.54 fT/ $\rm{Hz^{1/2}}$ was obtained by suppressing atomic spin exchange relaxation in the pulsed mode.  S. G. Li~et~al.achieved a sensitivity of 0.2~pT/$\rm{Hz^{1/2}}$ by adding two highly reflective mirrors to the vapor cell to reflect the beam through the cell several times$^{20}$. Dumke~et~al. demonstrated a cavity\mbox{-}enhanced all\mbox{-}optical  atomic magnetometer by placing a cesium vapor cell in a low finesse cavity$^{26}$.

For atomic magnetometer, the~photon shot noise(PSN) is one of the main limiting factors for the sensitivity. A~desirable method to counteract this limiting factor is to use a squeezed light, which is beyond PSN, to~replace the coherent light. Some research groups used different methods to generate squeezed light and have applied them to atomic magnetometers; all of these methods have successfully demonstrated the quantum enhancement effect of squeezed light on atomic magnetometers$^{27-29}$. In~2010, M. Mitchell's group preliminarily demonstrated the quantum enhancement measurement of rubidium atomic magnetometer by using 795 nm polarization-squeezed light with a squeezed level of -3.2 dB. The~sensitivity was improved from 46~nT/$\rm{Hz^{1/2}}$ to 32~nT/$\rm{Hz^{1/2}}$ $^{30}$. In~2021, this research group further introduced polarization-squeezed light into the Bell--Bloom all-optical atomic magnetometer, and~they also analyzed the main noise in different frequency segments~$^{31}$. Our group demonstrated a~rubidium atomic magnetometer based on the Faraday rotation and introduced Stokes operator $\widehat{S}_2$ polarization-squeezed light instead of coherent light; the quantum enhancement effect of the squeezed light was demonstrated at the analysis frequency of 10 kHz by using polarization-squeezed light with a squeezed level of $-$3.7~dB$^{32}$.  

In this paper, we describe how we built FID rubidium atomic magnetometer, and~separated the pump beam, the~$\pi / 2$ pulse of the RF magnetic field, and~the probe beam in time to avoid mutual effect caused by the time overlap of the three fields. We analyzed and optimized parameters of FID rubidium atomic magnetometer and demonstrated the single-pass and the triple-pass probe beam to explore effect of the optical path of interaction between probe beam and atomic ensemble without changing the spatial resolution, which provided a solid basis for further exploring improvement of the sensitivity of atomic magnetometer by employing Stokes operator $\widehat{S}_2$ polarization-squeezed~light.

\section{Experimental~Setup}
Our experimental setup is depicted in fig.1. A~20 $\times$ 20 $\times$ 20 $\rm{mm^{3}}$ naturally abundant rubidium vapor cell with a 795 nm anti\mbox{-}reflection coating on the outer surfaces of four windows, which also contained 100 Torr $N_{2}$ gas as a fluorescence-quenching gas to suppress the spin\mbox{-}polarization destruction caused by fluorescence with the random polarization state when the excited atoms fell back to the ground state; $N_{2}$ gas can also serves as a buffer gas to decrease the spin relaxation rate, which is caused by spin-exchange collisions between atoms and spin\mbox{-}destruction collisions between atoms and the vapor cell inner wall. The square flexible film electric heater with holes (with apertures$\sim$12 mm), which was driven by AC current, was used as heater. A~PT100 thermistor was used as a temperature sensor element, and~a commercial temperature control (ANTHONE LU\mbox{-}920) was used for temperature control. Temperature fluctuation of the heating system was not more than 1 $^{\circ}$C when the temperature was controlled at $\sim$100 $^{\circ}$C. The atomic vapor cell was placed in a four\mbox{-}layer permalloy ($\mu$\mbox{-}metal) magnetic shield to screen environmental magnetic field. The~shielding factor was more than 50,000 and the remanence was less than 1 nT. Meanwhile, two pairs of Helmholtz coils were placed along the Z\mbox{-}axis and X\mbox{-}axis, respectively, to generate static magnetic field \textbf{${B_{0}}$} and RF magnetic field \textbf{${B_{RF}}$}. The external\mbox{-}cavity diode laser (ECDL) of 795 nm provides a pump beam across two acousto\mbox{-}optical modulators (AOMs). The first AOM was mainly used for feedback to stabilize pump power and reduce impact of intensity fluctuations. The frequency shift generated by the first AOM was compensated by the second one, which was also used for switch control of the pump beam. The pump beam then became circularly polarized by a $\lambda$/4 wave plate after beam expansion, which entered the magnetic shield from the side hole, and travelled across the vapor cell via a rectangular prism. The beam diameter after beam expansion was about 10 mm, and the propagation direction passing through the vapor cell was consistent with the static magnetic field \textbf{${B_{0}}$}. Similarly, the~linearly polarized probe beam also crossed an AOM for the switch control. As shown in fig.1, we demonstrated the multi\mbox{-}pass measurement by adding two highly reflective mirrors to the probe beam path. The beam diameter was about 2 mm and the direction was roughly perpendicular to the RF magnetic field; it entered the~ polarimeter (composed of $\lambda$/2, a~Wollaston prism, and~a balanced differential detector) after passing through the vapor cell along the Y\mbox{-}axis in order to detect the rotation signal of the polarization plane of the probe beam. The data acquisition card DAQ (NI\mbox{-}USB6363) was used to collect the FID signal.

\begin{figure}
\includegraphics[width=8.5cm]{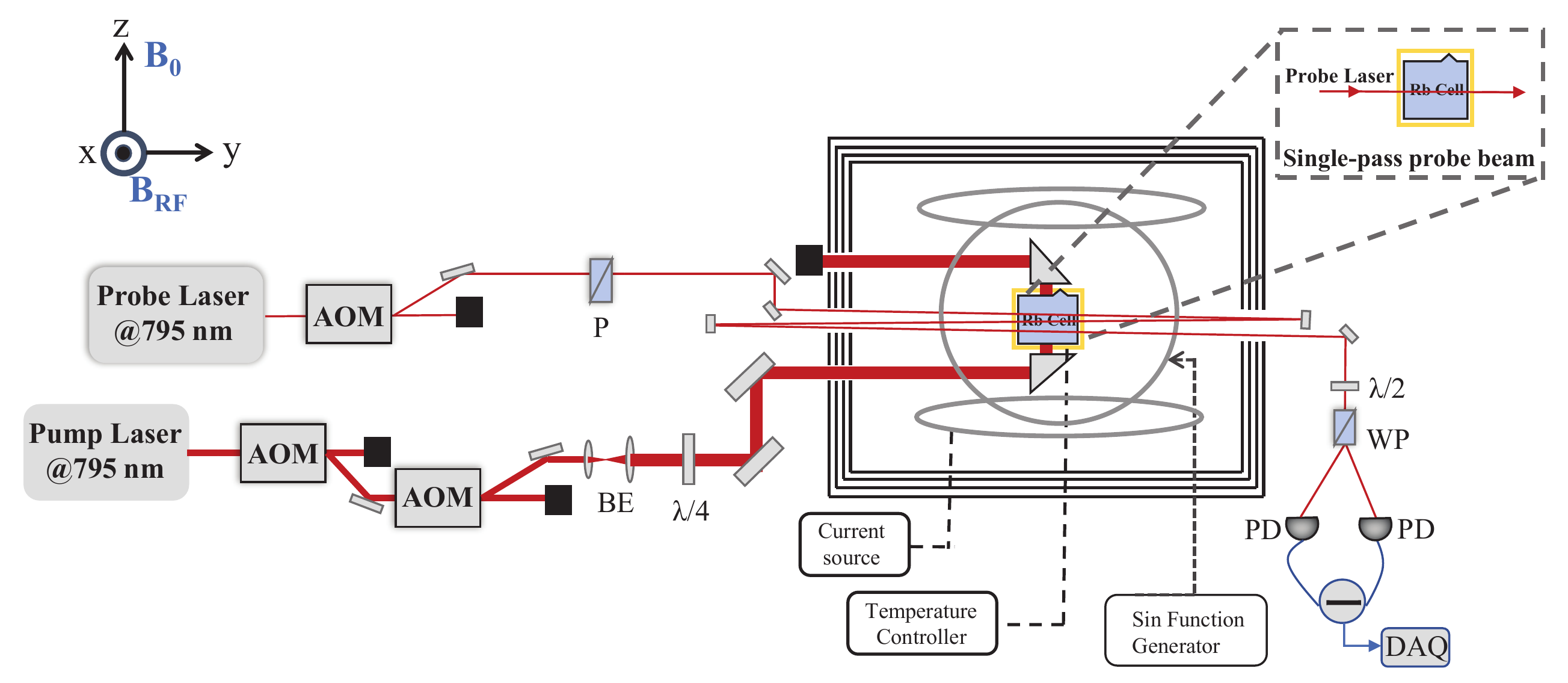}
\caption{Experimental setup. AOM: acoustic\mbox{-}optical modulator; BE: beam expander; $\lambda$/4: quarter\mbox{-}wave plate; $\lambda$/2: half\mbox{-}wave plate; P: polarizer; WP: Wollaston prism; PD: photoelectric detector; DAQ: data~acquisition.\label{fig1}}
\end{figure}

\section{Theoretical~Analysis }
Physics picture of spin polarization of rubidium atoms with 100 Torr $N_{2}$ gas pressure is shown in fig.2a. The~Zeeman energy levels of the rubidium $5^{2}S_{1/2}$ state were $m_{J}$ = +1/2 and $m_{J}$ = $-$1/2, respectively, without~considering the effect of the nuclear spin. When the circularly-polarized beam was resonant with the rubidium D1 line ($5^{2}S_{1/2}$--$5^{2}P_{1/2}$), atoms in the $m_{J}$ = $-$1/2 may be transferred to $m_{J}$ = +1/2, and~the atoms in the $m_{J}$ = +1/2 state did not absorb the circularly-polarized beam and were in the dark state. The~atomic spontaneous emission from the excited state fell back to the ground states $m_{J}$ = $-$1/2 and $m_{J}$ = +1/2. As~the process continued, most of the atoms were finally transferred to the dark state $m_{J}$ = +1/2, which was in preparation of the spin\mbox{-}polarized state.  In~this process, the~photons with random polarization were absorbed by $N_{2}$ gas to avoid the destruction of the atomic spin polarization$^{33}$. Due to the static magnetic field 
\textbf{${B_{0}}$}, $m_{J}$ = +1/2 and $m_{J}$ = $-$1/2 were Zeeman split at the Larmor frequency.

\begin{figure}
\includegraphics[width=8.5cm]{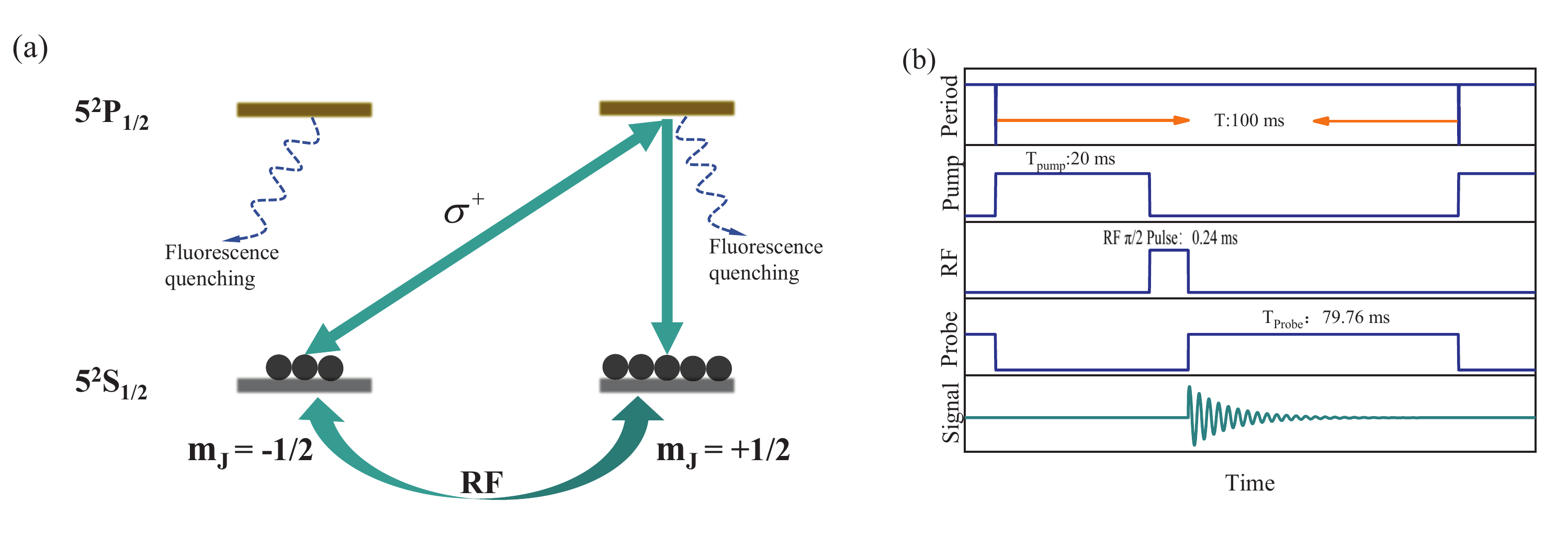}
\caption{(\textbf{a}) Physics picture of spin polarization rubidium ensemble with proper $N_2$ gas pressure; (\textbf{b})~Time sequence diagram and typical time~intervals.\label{fig2}}
\end{figure}

RF magnetic field with a Larmor frequency was applied after the atoms were populated to the spin\mbox{-}polarized state, the~atoms would back and forth between the ground states $m_{J}$ = +1/2 and $m_{J}$ = $-$1/2, which is the magnetic resonance transition. This process can also be represented by a Bloch sphere. The~Zeeman states $m_{J}$ = +1/2 and \mbox{$m_{J}$ = $-$1/2} can be regarded as stationary two-level system, which is expressed as $\vert{0}\rangle$ state and $\vert{1}\rangle$ state, corresponding to the north and south poles of the Bloch sphere, and~any point on the sphere represents the superposition of the two states. The~initial atomic spin polarization was in $\vert{0}\rangle$ state parallel to the direction of the static magnetic field $B_{0}$. The~atomic spin polarization can be rotated to any point on the Bloch sphere where the pulse duration of the RF magnetic field was different, which manifested as the different projection values of the spin\mbox{-}polarization vectors in the direction of the probe light. It then continued to evolve freely with a Larmor frequency around the static magnetic field \textbf{${B_{0}}$} until the thermal equilibrium state was reached. Typically, when the RF magnetic field was the $\pi / 2$ pulse, the~atomic spin polarization was rotated 90\textdegree\space from the north pole to the equatorial plane, and~the projection value of the spin-polarization vector was the largest in the probe direction. When the RF magnetic field was the $\pi$ pulse, the~atomic spin polarization was rotated 180\textdegree\space from the north pole to the south pole, and~the spin-polarization direction was parallel to the static magnetic field in which the projection value in the probe direction was the smallest. In~addition, the~strength of the RF magnetic field determined the interaction time between the RF magnetic field and atomic ensemble, and~the two were inversely proportional. However, as~the RF magnetic field strength increased, the~inhomogeneity of the magnetic field also increased, which might have caused other problems, such as the magnetic resonance linewidth broadening, hence it was indispensable to choose an appropriate RF magnetic field~strength.

This process could also be explained by a classical picture. In~the laboratory coordinate system, if~there is an atomic magnetic moment \textbf{$\mu$} procession around the magnetic field \textbf{${B_{0}}$} with an angular frequency of \textbf{$\omega_0$} due to torque \textbf{$L$} ($\textbf{$L$}= \textbf{$\mu$} \times \textbf{${B_{0}}$}$), then the dynamic equation of angular momentum \textbf{$F$} is as follows:

\begin{equation}
 d\textbf{$F$}/dt = \textbf{$\omega_0$} \times \textbf{$F$}  = \gamma \textbf{$F$} \times \textbf{$B_{0}$} 
\end{equation}

$\gamma$ is the gyromagnetic ratio  of the~atoms,. It can be ascertained from the above equation that $\textbf{$B_{0}$} = \textbf{$\omega$} / \gamma $, if~an RF magnetic field \textbf{$B_{RF}$}, whose angular frequency is \textbf{$\omega$}, is applied. At~this time, if~the rotating coordinate system is introduced, then the procession equation of angular momentum is as follows:
\begin{equation}
    d\textbf{$F$}/dt = \gamma \textbf{$F$} \times \textbf{$B_{eff}$}        
\end{equation}

where $\textbf{$B_{eff}$} = \textbf{$B_{0}$} + \textbf{$\omega_0$} / \gamma + \textbf{$B_{RF}$}$ is the effective magnetic field. In~this experiment, we investigated the variation in the magnetic moment in the case of $\omega = \omega_0$. The~projection vector of the atomic magnetic moment in the direction of the probe beam could be detected due to the Faraday rotation effect, and~the output signal was proportional to the projection value$^{34}$. In~an atomic vapor cell, collisions (between atoms, between~atoms and the inner walls of the vapor cell, and~between atoms and the buffer gas) result in a change in the orientation of the magnetic moments. If~the relaxation process of the atomic polarization is considered, and~the initial phase of the atomic magnetic moment is assumed to be $\varphi$, FID signal output by the polarimeter can be expressed as follows$^{35}$:
\begin{equation}
    S_{signal} = A \mu sin(\gamma B_{RF} t)cos(\omega_0 t +\varphi) \times exp(-t/T_{2})
\end{equation}
where \textit{A} is the proportionality coefficient and $T_2$ is the atomic spin transverse relaxation time. According to this equation, we know that FID signal is related to the RF magnetic field strength. When $\gamma B_{RF} t = \pi / 2$, that is, the~RF magnetic field operates as $\pi / 2$ pulse, the~amplitude of the output signal is maximum. The~signal-to-noise ratio (SNR) and the full width at half maximum  (FWHM) $\Gamma$ after a fast Fourier transform (FFT) of the output signal, directly affect the sensitivity of the magnetometer$^{33}$:
\begin{equation}
        \delta B = \hbar \Gamma / \left[ g \mu_B \times  (SNR) \right]
\end{equation}
Here $\hbar$ is the reduced Planck constant, \textit{g} is the Lande factor, $\mu_B$ is the Bohr~magneton.

Time sequence diagram is shown in fig.2b. In~a complete period T, we first turned the pump beam on long enough to ensure that atomic spin was sufficiently polarized. The~pump beam was then switched off and the RF magnetic field was turned on  for $\pi / 2$ pulse. Finally, the~probe beam was turned on for FID signal detection. This design could effectively prevent the mutual effect caused by the time overlap of the three fields.

The mode of setting the RF magnetic field ($\pi / 2$ pulse mode) is shown in fig.3a. With~the help of the RF magnetic field, the~atoms were made to transition back and forth between the adjacent Zeeman energy levels. Accordingly, the~polarimeter output a gradually decaying signal, similar to the breathing mode. The~RF magnetic field was turned off at the maximum value of the first breathing mode, which was the $\pi / 2$ pulse. Significantly, when the RF magnetic field strength increased, the~time to complete a breathing mode became shorter, and~the time of the $\pi/2$ pulse was also shorter. Therefore, if~the RF magnetic field strength was changed, the~opening time of the RF magnetic field should be changed simultaneously to maintain the $\pi / 2$ pulse. When the RF magnetic field was turned off, the~probe beam (\mbox{$T_{probe}$ = T $-$ $T_{pump}$ $-$ RF $\pi / 2$ Pulse}) was turned on to detect the Larmor precession frequency, as~shown in fig.3b. A~typical FWHM after FFT is about 332 Hz, as~shown in fig.3c. The~typical parameter T was set at 100 ms, $T_{pump}$ was set at 20 ms, and~the RF magnetic field was set as $\pi / 2$ pulse mode in the following study of optimal~parameters.

\begin{figure}
\includegraphics[width=8.5cm]{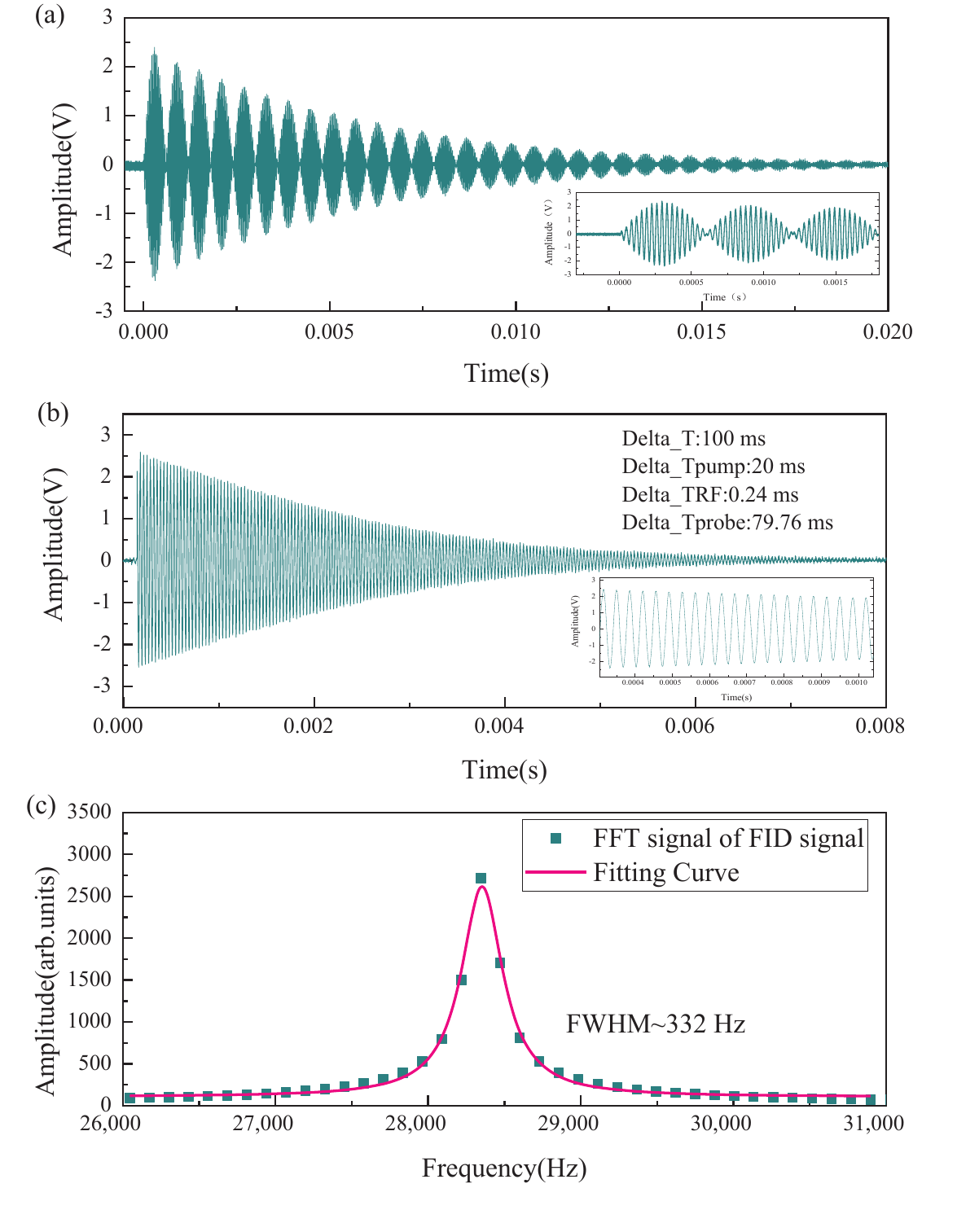}
\caption{(\textbf{a}) Magnetic resonance signal with RF magnetic field; (\textbf{b}) FID signal; (\textbf{c}) FFT of the FID~signal.\label{fig3}}
\end{figure}

\section{Parameter~Optimization}

\subsection{Rubidium Atomic Number~Density}
When we studied influence of the atomic number density of  $\rm{^{85}Rb}$ and  $\rm{^{87}Rb}$ on the FID signal amplitude and the FWHM after FFT, we kept the other parameters as follows: the coil current that generated the RF magnetic field was 4 mA; the probe power was 100~${\mu}$W; and the frequency was red detuned to 45 GHz relative to the 5$S_{1/2}$(\mbox{$F_{g}$ = 3})--5$P_{1/2}$(\mbox{$F_{g}$ = 2}) transition of $\rm{^{85}Rb}$ and the 5$S_{1/2}$($F_{g}$ = 2)--5$P_{1/2}$($F_{g}$ = 1) transition of $\rm{^{87}Rb}$, respectively. The~pump power was 15 mW, and~it was resonant with the 5$S_{1/2}$($F_{g}$ = 3)--5$P_{1/2}$($F_{g}$ = 2) transition of  $\rm{^{85}Rb}$ and the 5$S_{1/2}$(\mbox{$F_{g}$ = 2})--5$P_{1/2}$($F_{g}$ = 1) transition of $\rm{^{87}Rb}$, respectively. For~our atomic vapor cell, increase in the atomic number density was not necessarily beneficial to the FWHM and the SNR. As~shown in fig.4a,b, with~the increase in the atomic number density, FWHM showed an increasing trend, and~SNR of $\rm{^{85}Rb}$ and  $\rm{^{87}Rb}$ reached the optimal value at 60 $^{\circ}$C. FWHM and the SNR of $\rm{^{87}Rb}$ were poor compared with  $\rm{^{85}Rb}$ due to the lower natural abundance, thus the~temperature was controlled at 60 $^{\circ}$C and the $\rm{^{85}Rb}$ was employed in subsequent~experiments.

\begin{figure}
\includegraphics[width=8.5cm]{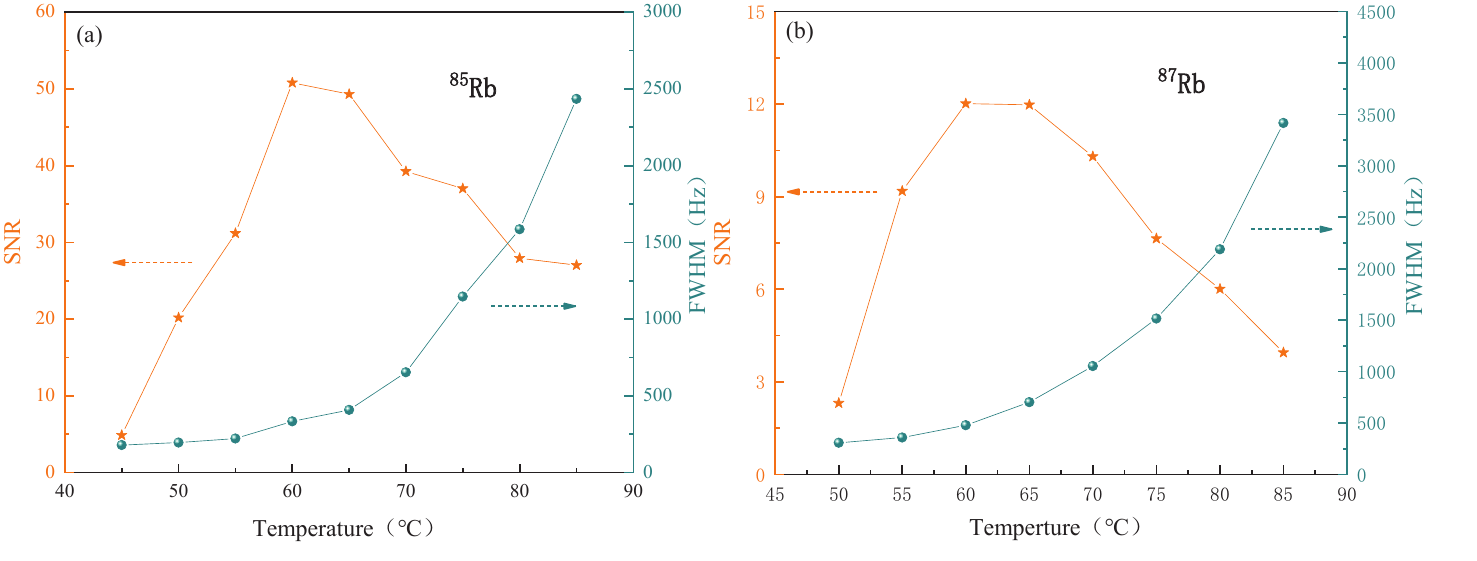}
\caption{Effect of different atomic number densities on FWHM and SNR after FFT of $\rm{^{85}Rb}$ atomic ensemble \space(\textbf{a}) and $\rm{^{87}Rb}$ atomic ensemble \space(\textbf{b}). \label{fig4}}
\end{figure}

\subsection{The Power of the Pump~Beam}
As shown in fig.5a, with~increase of the pump power, the~signal amplitude increased, while FWHM decreased. When the pump power reached 4 mW, the~atomic polarization reached its maximum value. As~the pump power increased further, the~signal amplitude and FWHM tended to be stable. We controlled the pump power at 15 mW in~subsequent experiments.

\begin{figure}
\includegraphics[width=8.5cm]{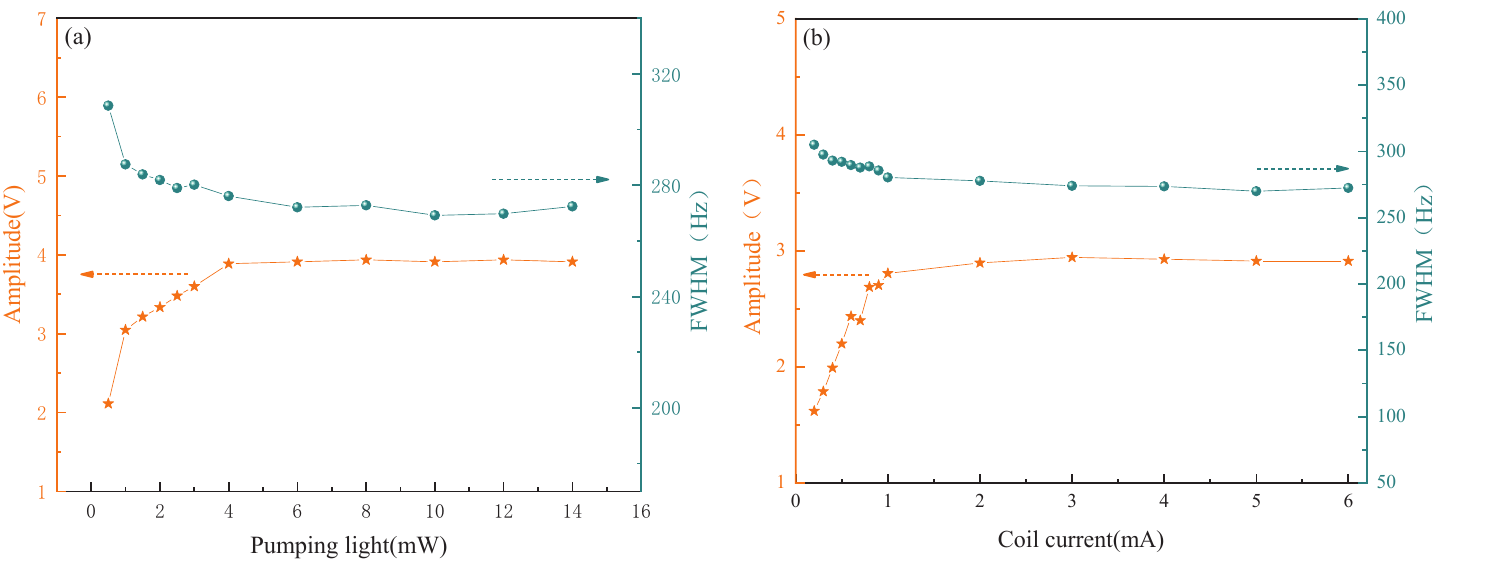}
\caption{(\textbf{a}) Dependence of signal amplitude and FWHM on pump power---coil current that generated RF magnetic field was 4 mA; (\textbf{b}) Dependence of signal amplitude and FWHM on RF magnetic field---pump power was 15 mW. For (\textbf{a},\textbf{b}), the~conditions for other parameters were as follows: probe power was 100 $\mu$W, and frequency was red detuned to 45 GHz relative to~5$S_{1/2}$($F_{g}$ = 3)--5$P_{1/2}$($F_{g}$ = 2) transition of $\rm{^{85}Rb}$ atoms. Pump beam was resonant with the 5$S_{1/2}$($F_{g}$ = 3)--5$P_{1/2}$($F_{g}$ = 2) transition of $\rm{^{85}Rb}$ atoms.\label{fig5}}
\end{figure}

\subsection{Different RF Magnetic Field~Strength}
Strength of the RF magnetic field also affected the signal amplitude and FWHM. We changed the RF magnetic field strength by changing the coil current, as~shown in fig.5b. When the coil current was small, only partially polarized atoms were prepared in a direction perpendicular to atomic polarization, so both the amplitude and FWHM of Faraday rotation signal were not optimal. When the coil current increased to 2~mA, it reached saturation. With further increase of the RF magnetic field strength, the~amplitude and FWHM of Faraday rotation signal tended to be stable, and~we did not observe any broadening caused by inhomogeneity of the RF magnetic field. We took 3 mA for subsequent~experiments.

\subsection{Different Static Magnetic Field~Strengths}
Fig.6 shows changes in the amplitude and FWHM of Faraday rotation signal under different static magnetic fields. With increase in the static magnetic field, the~signal amplitude did not change significantly, but~FWHM increased significantly when the static magnetic field reached a certain value. This could be attributed to the nonlinear Zeeman effect and inhomogeneity caused by increase of the static magnetic~field. 
\begin{figure}
\includegraphics[width=7cm]{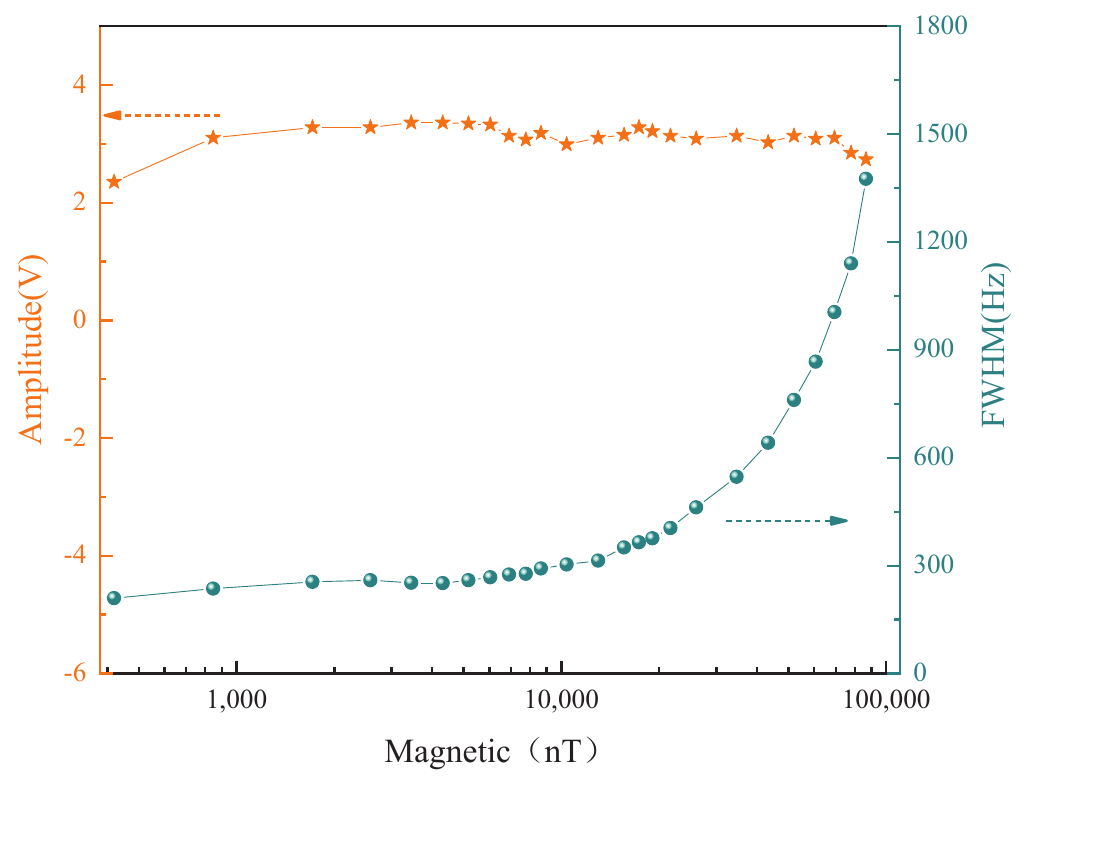}
\caption{Dependence of FID signal amplitude and FWHM on static magnetic field.\label{fig6}}
\end{figure}

\section{Performance Comparison of the Magnetometer under a Single-Pass and a Triple-Pass Probe~Beam}

Sensitivity of the magnetometer was compared and studied in the case of the single-pass and the triple-pass probe beam. In~fig.7a, we demonstrated typical signals collected by DAQ when the period was about 7.7 ms and the pump beam was turned on for 5 ms. The~inset represents the signal within a period. The~output signal was then processed by FFT, and~the peak value was fitted. The~fitted center frequency was the Larmor frequency, as~shown in fig.7b,c. It could be seen compared with the single-pass probe beam case, the~signal amplitude increased significantly in the case of the triple-pass probe beam due to the longer optical path of the interaction between probe beam and atomic ensemble.

\begin{figure}
\includegraphics[width=8.5cm]{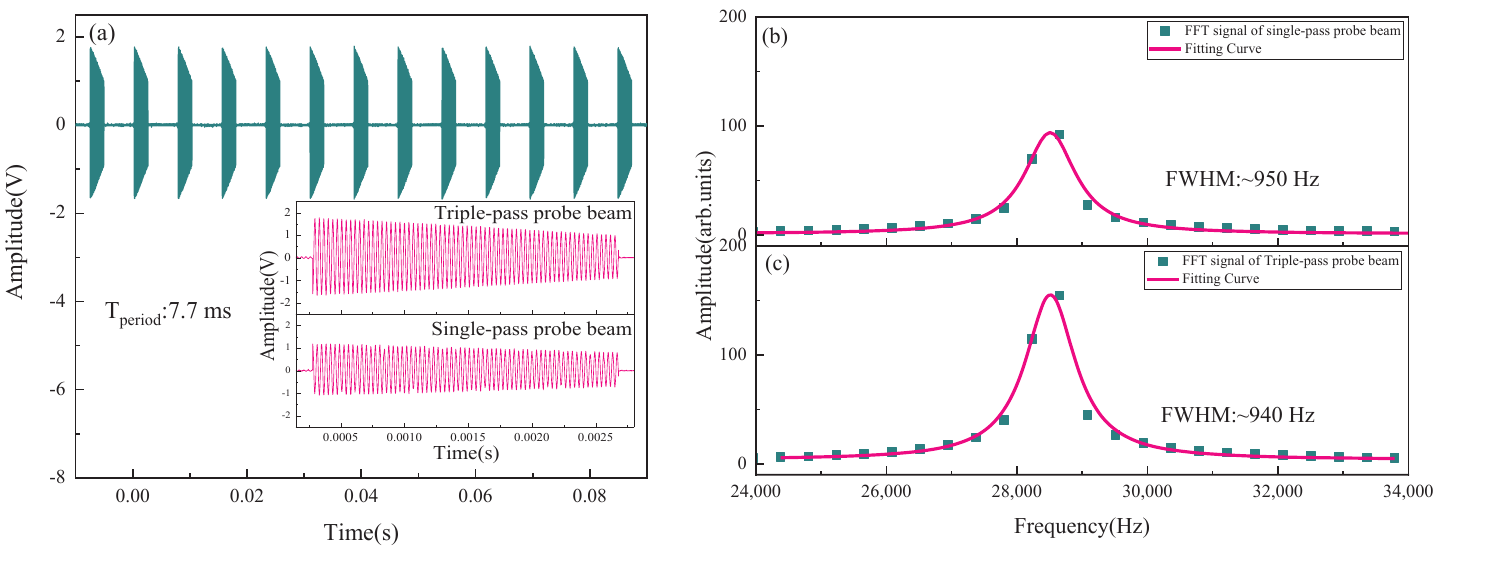}
\caption{(\textbf{a}) Output signal of the balanced differential photodiodes (insert: FID signals for the single-pass and the triple-pass probe beam cases in the same period); (\textbf{b},\textbf{c}), FFT of FID signals for the single-pass and the triple-pass probe beam cases in one~period. \label{fig7}}
\end{figure}

In fig.7b,c, Larmor frequency values were converted to the magnetic field values by the formula $\omega = \gamma \cdot$ B (the gyromagnetic ratio $\gamma$ of ground state (F = 3) of $\rm{^{85}Rb}$ is 4.66743 Hz/nT). A~series of the magnetic field values were obtained by repeated measurements, and~the power spectral density (PSD) was calculated. Fig.8a--d were the magnetic field values and sensitivity of the magnetometer for the single-pass and the triple-pass probe beam cases, with~a sampling rate of 130 Hz and a sampling period number of 6478, respectively. The~sensitivity of the magnetometer was 21.2 pT/$\rm{Hz^{1/2}}$ and 13.4~pT/$\rm{Hz^{1/2}}$, respectively, which was calculated by statistically averaging sensitivity values, indicated that the sensitivity had been significantly improved. According to statistical average of the magnetic field value distribution, the~value of the static magnetic field was 5.98450(2) ± (4) ${\mu}$T for the single-pass probe beam case and 5.98331(2) ± (2) ${\mu}$T for the triple-pass probe beam case. In~other words, error distribution of the magnetic field value did not increase with the increase in the optical path of the interaction between probe beam and atomic ensemble.
\begin{figure}
\includegraphics[width=8.5cm]{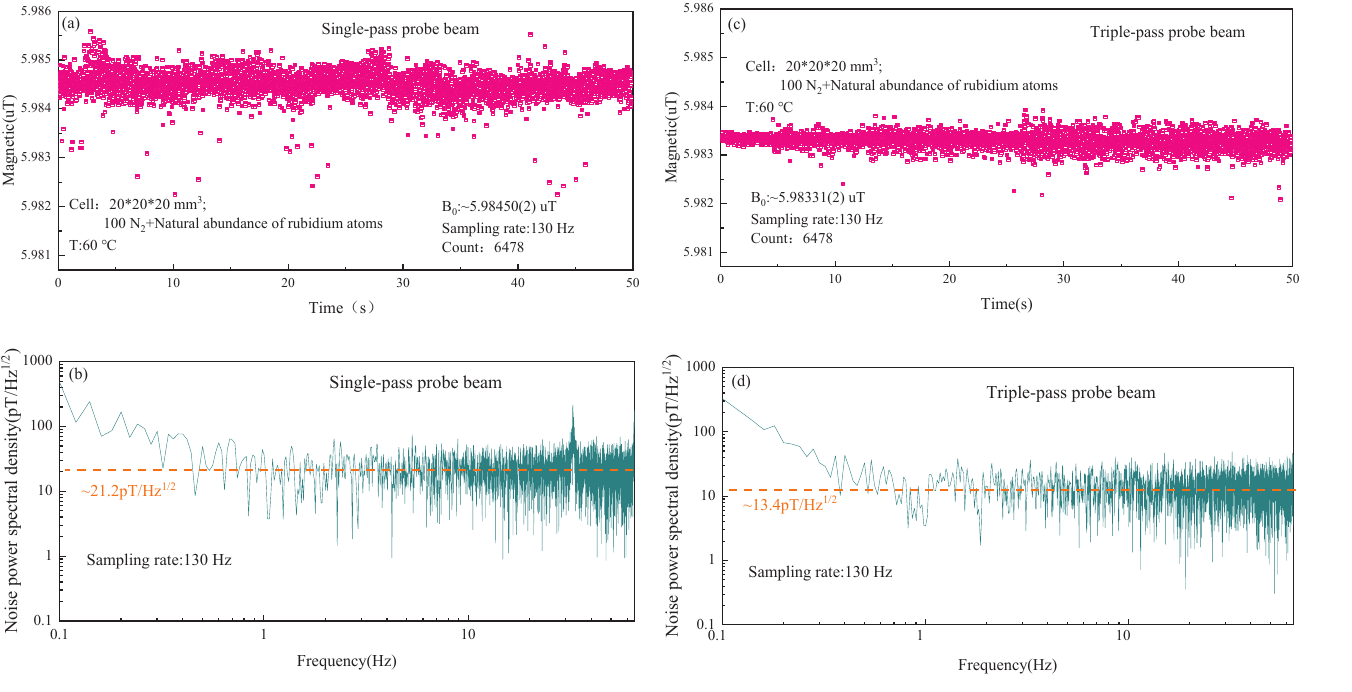}
\caption{(\textbf{a}) Measured data set of the magnetic field for the single-pass probe beam case for a period of $\sim$7.7 ms; (\textbf{b})~Calculated PSD of data in  (\textbf{a})---the dotted orange line indicates the calculated noise floor $\sim$21.2~pT/$\rm{Hz^{1/2}}$ with a bandwidth of 65 Hz; (\textbf{c}) Measured data set of the magnetic field for the triple-pass probe beam case for a period of $\sim$7.7 ms; (\textbf{d}) Calculated PSD of data in (\textbf{c})---the dotted orange line indicates the calculated noise floor $\sim$13.4 pT/$\rm{Hz^{1/2}}$ with a bandwidth of 65~Hz. \label{fig8}}
\end{figure}

\begin{table*}
    \centering
    \caption{Parameters for the single\mbox{-}pass and the triple\mbox{-}pass probe beam cases of FID magnetometers.}
    \begin{tabular}{ccccc}
    \hline
      & Signal (V) & Noise(V) &  FWHM(Hz) & Sensitivity (pT/{$\rm{Hz^{1/2}}$})  \\
     \hline
Single-pass probe case	& 0.864	& 0.056 & 950 & 21.2\\
Triple-pass probe case	& 1.368	& 0.056 & 940 & 13.4\\
    \hline
    \end{tabular}
\end{table*}

Parameters for the single-pass and the triple-pass probe beam cases are shown in Tab.1. When the probe beam triply passed vapor cell, the~signal amplitude increased, but~the noise amplitude and FWHM were almost unchanged. It could be seen from the table that the sensitivity of the magnetometer could be further improved by continuously increasing the optical path of the interaction, and FWHM would not further widen, but~it should be noted that the probe light needs to pass through the vapor cell more times, which requires a higher homogeneity in the magnetic~field.

\section{Sensitivity Analysis and~Discussion}
For optically\mbox{-}pumped atomic magnetometer, homogeneity of the magnetic field was also the main factor of the sensitivity. The~static magnetic field ${B_{0}}$ and the RF magnetic field ${B_{RF}}$ used in our experimental system also had a potential risk of inhomogeneity. In~particular, increase in the magnetic field inhomogeneity not only broadens the magnetic resonance spectrum but also has a negative impact on sensitivity. This is the direction of our further research. When the system was optimized, sensitivity of the magnetometer was mainly limited by quantum noise, that is, atomic spin projection noise $\delta{B_{at}}$ and PSN $\delta{B_{ph}}$ $^{36-38}$. Atomic spin projection noise can be suppressed by atomic spin squeezing$^{39,40}$. The~PSN can be suppressed by introducing the squeezed light. In~previous work, our group demonstrated the improvement of sensitivity by introducing a Stokes operator $\hat{S}_2$ polarization\mbox{-}squeezed light in the rubidium atomic magnetometer based on the Faraday rotation$^{32}$, and~we also systematically analyzed the effect of frequency detuning on the squeezed level; we found that there was an obvious loss in the squeezed level when the probe light is resonant with rubidium atomic atoms. This indicated that the probe light with large frequency detuning was a better choice for compatibility with the polarization-squeezed light. In the Mz magnetometer, which we demonstrated, there was only a circularly polarized light resonated with the atomic transition line and it was not very compatible with the squeezed light$^{41}$. In FID type rubidium atomic magnetometer system, the~probe light with large frequency detuning fully met the introduction conditions of the polarization-squeezed light; however, in~our system, optimization to make the system noise reach the PSN before the introduction of squeezed light was crucially important, such as the intensity fluctuation of the light, the~inhomogeneity of the magnetic field, and etc. Additionally, to~avoid the potential risk of inhomogeneity of the RF magnetic field on the sensitivity, as~well as to reduce the additional time added by the RF magnetic field, the~Bell--Bloom scheme with the pump light amplitude modulation instead of the RF magnetic field could be considered in subsequent experiments, while a Stokes operator $\hat{S}_2$ polarization\mbox{-}squeezed light was employed to beyond the PSN.

\section{Conclusions}
 In this paper, we described physics picture of FID type optically\mbox{-}pumped rubidium magnetometer in detail. Dependence relationship between the parameters and FID signal was also systematically analyzed and optimized. Importantly, the~pump beam, $\pi/2$ pulse of the RF magnetic field, and~the probe beam in our system were sequentially controlled to be independent of each other in order to eliminate crosstalk between the three fields. The~influence of the optical path between probe beam and atomic ensemble on sensitivity was demonstrated. It was observed that interaction of the optical path was proportional to SNR of FID magnetometer and sensitivity was improved from 21.2 pT/$\rm{Hz^{1/2}}$ to 13.4 pT/$\rm{Hz^{1/2}}$, without changing the spatial resolution. Additionally, the~dual\mbox{-}beam FID rubidium atomic magnetometer with large detuning of the probe beam frequency avoided atomic resonance absorption at the detection phase, which was caused by a measurement deviation$^{42}$. This result laid a foundation for subsequent employing Stokes operator $\widehat{S}_2$ polarization\mbox{-}squeezed light$^{2}$ to improve SNR, and~further improve the sensitivity of the magnetometer. Compared to previous works$^{32,41}$, FID type rubidium atomic magnetometer more broadly and accurately measured magnetic fields, and~was not limited to tracking magnetic field in real-time. In~addition, we preliminarily demonstrated quantum enhancement in a single\mbox{-}beam linearly polarized Faraday rotation magnetometer, via 795 nm Stokes operator $\widehat{S}_2$ polarization\mbox{-}squeezed light; the sensitivity was improved from 28.3 pT/$\rm{Hz^{1/2}}$ to 19.5 pT/$\rm{Hz^{1/2}}$$^{32}$. Moreover, in~Ref.[31], they achieved a $17\%$ quantum enhancement of magnetometer sensitivity by employing a polarization\mbox{-}squeezed light. This means that we could make SNR, measured by the polarization rotation angle of the probe light, greater than PSN and sensitivity could be improved significantly via introducing a Stokes operator $\hat{S}_2$ polarization\mbox{-}squeezed light, instead of coherent light.

\section*{Funding}

This project was supported by the National Natural Science Foundation of China (Grant Nos. 11974226, 61875111) and  the Shanxi Provincial Graduate Education Innovation Project (Grant No.2022Y022).

\section*{Conflict of Interest}

The authors have no conflicts to disclose.

\nocite{*}
\bibliography{aipsamp}

\section*{References}

$^1$ Casola, F.; van der Sar, T.; Yacoby, A. Probing condensed matter physics with magnetometry based on nitrogen-vacancy centres in diamond. {\em Nat. Rev. Mater.} {\bf 2018}, {\em 3}, 17088.\\
$^2$ Vasyukov, D.; Anahory, Y.; Embon, L.; Halbertal, D.; Cuppens, J.; Neeman, L.; Finkler, A.; Segev, Y.; Myasoedov, Y.; Rappaport, M.L.; et al. A Scanning Superconducting  Quantum Interference  Device Single Electron Spin Sensitivity. {\em Nat. Nanotechnol.} {\bf 2013}, {\em 8}, 639--644.\\
$^3$ Sander, T.H.; Preusser, J.; Mhaskar, R.; Kitching, J.; Trahms, L.; Knappe, S. Magnetoencephalography with a chip-scale atomic magnetometer. {\em Biomed. Opt. Express} {\bf 2012}, {\em 3}, 981--990.\\
$^4$ Kamada, K.; Sato, D.; Ito, Y.; Natsukawa, H.; Okano, K.; Mizutani, N.; Kobayashi, T. Human magnetoencephalogram measurements using newly developed compact module of high-sensitivity atomic magnetometer. {\em Jpn. J. Appl. Phys.} {\bf 2015}, {\em 54},~026601.\\
$^5$ Alem, O.; Sander, T.H.; Mhaskar, R.; LeBlanc, J.; Eswaran, H.; Steinhoff, U.; Okada, Y.; Kitching, J.; Trahms, L.; Knappe, S. Fetal magnetocardiography measurements with an array of microfabricated optically pumped magnetometers. {\em Phys. Med. Biol.} {\bf 2015}, {\em 60}, 4797--4811.
$^6$ Mateos, I.; Patton, B.; Zhivun, E.; Budker, D.; Wurm, D.; Ramos-Castro, J. Noise characterization of an atomic magnetometer at sub-millihertz frequencies. {\em Sens. Actuat. A Phys.} {\bf 2015}, {\em 224}, 147--155.\\
$^7$ Korth, H.; Strohbehn, K.; Tejada, F.; Andreou, A.G.; Kitching, J.; Knappe, S.; Lehtonen, S.J.; London, S.M.; Kafel, M. Miniature atomic scalar magnetometer for space based on the rubidium isotope $^{87}$Rb. {\em J. Geophys. Res. Space Phys.} {\bf 2016}, {\em 121}, 7870--7880.\\
$^8$ Jiang, M.; Su, H.W.; Garcon, A.; Peng, X.H.; Budker, D. Search for axion-like dark matter with spin-based amplifiers. {\em Nat. Phys.} {\bf 2021}, {\em 17}, 1402--1407.\\
$^9$ Paoletti, V.; Buggi, A.; Pasteka, R. UXO Detection by Multiscale Potential Field Methods. {\em Pure Appl. Geophys.} {\bf 2019}, {\em 176}, 4363--4381.\\
$^{10}$ Shockley, J.A.; Raquet, J. Navigation of ground vehicles using magnetic field variations. {\em Navigation} {\bf 2015}, {\em 61}, 237--252.\\
$^{11}$ Canciani, A.; Raquet, J. Airborne magnetic anomaly navigation. {\em IEEE Trans. Aerosp. Electron. Syst.} {\bf 2017}, {\em 53}, 67--80.\\
$^{12}$ Allred, J.; Lyman, R.N.; Kornack, T.W.; Romalis, M.V. High-sensitivity atomic magnetometer unaffected by spin-exchange relaxation. {\em Phys. Rev. Lett.} {\bf 2002}, {\em 89}, 130801.\\
$^{13}$ Gu, Y.; Sh, R.H.; Wang, Y.H. Study on sensitivity-related parameters of distributed feedback laser-pumped cesium atomic magnetometer. {\em Acta Phys. Sin.} {\bf 2014}, {\em 63}, 110701. (in Chinese).\\
$^{14}$ Su, S.; Zhang, G.; Bi, X.; He, X.; Zheng, W.; Lin, Q. Elliptically polarized laser-pumped Mx magnetometer towards applications at room temperature.
{\em Opt. Express} {\bf 2019}, {\em 27}, 33027.\\
$^{15}$ Belfi, J.; Bevilacqua, G.; Biancalana, V.; Cartaleva, S.; Dancheva, Y.; Moi, L. Cesium coherent population trapping magnetometer for cardiosignal detection in an unshielded environment. {\em J. Opt. Soc. Am. B} {\bf 2007}, {\em 24}, 2357--2362.\\
$^{16}$ Bell, W.E.; Bloom, A.L. Optically driven spin precession. {\em Phys. Rev. Lett.} {\bf 1961}, {\em 6}, 280--281.\\
$^{17}$ Savukov, I.M.; Romalis, M.V. NMR detection with an atomic magnetometer. {\em Phys. Rev. Lett.} {\bf 2005}, {\em 94}, 123001.
$^{18}$ Zhang, R.; Klinger, E.; Bustos, F.P.; Akulshin, A.; Guo, H.; Wickenbrock, A.; Budker, D. Stand-Off Magnetometry with Directional Emission from Sodium Vapors. {\em Phys. Rev. Lett.} {\bf 2021}, {\em 127}, 173605.
$^{19}$ Hunter, D.; Piccolomo, S.; Pritchard, J.D.; Brockie, N.L.; Dyer, T.E.; Riis, E. Free-induction-decay magnetometer based on a~microfabricated Cs vapor cell. {\em Phys. Rev. A} {\bf 2018}, {\em 10}, 014002.\\
$^{20}$ Li, S.G.; Liu, J.S.; Jin, M.; Tetteh, A.K.; Dai, P.F.; Xu, Z.K.; Eric-Theophilus, N.T. A kilohertz bandwidth and sensitive scalar atomic magnetometer using an optical multipass cell. {\em Measurement} {\bf 2022}, {\em 190}, 110704.\\
$^{21}$ Liu, C.; Dong, H.F.; Sang, J.J. Sub-millimeter-resolution magnetic field imaging with digital micromirror device and atomic vapor cell. {\em Appl. Phys. Lett.} {\bf 2021}, {\em 119}, 114002.\\
$^{22}$ Hunter, D.; Jiménez-Martínez, R.; Herbsommer, J.; Ramaswamy, S.; Li, W.; Riis, E. Waveform reconstruction with a Cs based free-induction-decay magnetometer. {\em Opt. Express} {\bf 2018}, {\em 26}, 30523--30531.\\
$^{23}$ Miao, P.X.; Yang, S.Y.; Wang, J.X.; Lian, J.Q.; Tu, J.H.; Yang, W.; Cui, J.Z. Rubidium atomic magnetometer based on pump-probe nonlinear magneto-optical rotation. {\em Acta Phys. Sin.} {\bf 2017}, {\em 66}, 160701. (in Chinese).\\
$^{24}$Wilson, N.; Perrella, C.; Anderson, R.; Luiten, A.; Light, P. Wide-bandwidth atomic magnetometry via instantaneous-phase retrieval. {\em Phys. Rev. Res.} {\bf 2020}, {\em 2}, 013213.\\
$^{25}$ Sheng, D.; Li, S.; Dural, N.; Romalis, M.V. Subfemtotesla Scalar Atomic Magnetometry Using Multipass Cells. {\em Phys. Rev. Lett.} {\bf 2013}, {\em 110}, 160802.\\
$^{26}$ Crepaz, H.; Ley, L.Y.; Dumke, R. Cavity enhanced atomic magnetometry. {\em Sci. Rep.} {\bf 2015}, {\em 5}, 15448.\\
$^{27}$ Horrom, T.; Singh, R.; Dowling, J.P.; Mikhailov, E.E. Quantum-enhanced magnetometer with low frequency squeezing. {\em Phys. Rev. A} {\bf 2012}, {\em 86}, 023803.\\
$^{28}$ Novikova, I.; Mikhailov, E.E.; Xiao, Y. Excess optical quantum noise in atomic sensors. {\em Phys. Rev. A} {\bf 2014}, {\em 91}, 051804.\\
$^{29}$ Otterstrom, N.; Pooser, R.C.; Lawrie, B.J. Nonlinear optical magnetometry with accessible in~situ optical squeezing. {\em Opt. Lett.} {\bf 2014}, {\em 39}, 1012001.\\
$^{30}$ Wolfgramm, F.; Cere, A.; Beduini, F.A.; Predojević, A.; Koschorreck, M.; Mitchell, M.W. Squeezed-Light Optical Magnetometry. {\em Phys. Rev. Lett.} {\bf 2010}, {\em 105}, 053601.\\
$^{31}$ Troullinou, C.;  Jiménez-Martínez, R.;  Kong, J.; Lucivero, V.G.; Mitchell, M.W. Squeezed-light enhancement and backaction evasion in a high sensitivity optically pumped magnetometer. {\em Phys. Rev. Lett.} {\bf 2021}, {\em 127}, 193601.\\
$^{32}$ Bai, L.L.; Wen, X.; Yang, Y.L.; Zhang, L.L.; He, J.; Wang, Y.H.; Wang, J.M. Quantum-Enhanced  Rubidium Atomic  Magnetometer Based Faraday Rotation  Via 795 Nm Stokes Operator Squeezed Light. {\em J. Opt.} {\bf 2021}, {\em 23}, 085202.\\
$^{33}$ Budker, D.; kimbal, D.F.J. \textit{Optical Magnetometry}; Cambridge University Press: Cambridge, UK, 2013; pp. 98--99.\\
$^{34}$ Miao, P.X.; Zheng, W.Q.; Yang, S.Y.; Wu, B.; Cheng, B.; Tu, J.H.; Ke, H.L.; Yang, W.; Wang, J.; Cui, J.Z.; et al. Wide-range and self-locking atomic magnetometer based on free spin precession. {\em J. Opt. Soc. Am. B} {\bf 2019}, {\em 36}, 819--828.\\
$^{35}$ Yang, B.; Miao, P.X.; Shi, Y.C.; Feng, H.; Zhang, J.M; Cui, J.Z; Liu, Z.D.  Theoretical and Experimental Studies on Classic Physical Picture of Two-Level Magnetic Resonance. {\em Chin. J. Lasers} {\bf 2020}, {\em 47}, 1012001. (in Chinese).\\
$^{36}$ Budker, D.; Romalis, M. Optical Magnetometry. {\em Nat. Phys.} {\bf 2007}, {\em 3}, 227--234.\\
$^{37}$ Budker, D.; Gawlik, W.; Kimball, D.F.; Rochester, S.M.; Yashchuk, V.V.; Weis, A. Resonant nonlinear magneto-optical effects in atoms. {\em Rev. Mod. Phys.} {\bf 2002}, {\em 74}, 1153--1201.\\
$^{38}$ Pustelny, S.; Wojciechowski, A.; Gring, M.;  Kotyrba, M.; Zachorowski, J.; Gawlik, W. Magnetometry Based Nonlinear Magneto-Optical Rotation Amplitude-Modul. {Light. J. Appl. Phys.} {\bf 2008}, {\em 103}, 063108.\\
$^{39}$ Bao, H.; Duan, J.; Jin, S.; Lu, X.; Li, P.; Qu, W.;  Wang, M.; Novikova, I.; Mikhailov, E.E.; Zhao, Ka.; et al. Spin squeezing of ${10^{11}}$ atoms by prediction and retrodiction measurements. {\em Nature} {\bf 2020}, {\em 581}, 159--163.\\
$^{40}$ Shah, V.; Vasilakis, G.; Romalis, M.V. High bandwidth atomic magnetometery with continuous quantum nondemolition measurements. {\em Phys. Rev. Lett.} {\bf 2010}, {\em 104}, 013604.\\
$^{41}$ Zhang, L.L.; Bai, L.L.; Yang, Y.L.;  Yang, Y.B.; Wang, Y.H.; Wen, X.; He, J.; Wang, J.M. Improving sensitivity of an optically pumped rubidium atomic magnetometer with a repumping laser. {\em Acta Phys. Sin.} {\bf 2021}, {\em 70}, 230702. (in Chinese).\\
$^{42}$ Hunter, D.; Dyer, T.E.; Riis, E. Accurate optically pumped magnetometer based on Ramsey-style interrogation. {\em Opt. Lett.} {\bf 2022}, {\em 47}, 1230--1233.\\

\end{document}